%% file: MARAMI2013.tex
\documentclass{article-hermes}

\usepackage{amssymb}
\usepackage{graphicx}
\usepackage{url}

\usepackage{subfigure}
\usepackage{enumerate}
\usepackage{MyAlgorithm}
\usepackage{MyAlgorithmic}
\usepackage{array}
\usepackage{multirow}
\usepackage{color}

\begin{document}



\journal{Conférence MARAMI}{}{}{2013}{1}{12}

\title[Dynamiques globales et locales]{Dynamiques globales et locales dans un réseau de télécommunications}
\subtitle{}

\author{Erick Stattner et Martine Collard}

\address{%
Laboratoire LAMIA\\
Université des Antilles et de la Guyane\\
FRANCE\\
\{estattne,mcollard\}@univ-ag.fr}

\resume{
	\input{section/abstractFr.tex}
	\\
}

\abstract{
	\input{section/abstractEn.tex}
	\\
}

\motscles{
Réseaux sociaux,
réseaux complexes,
évolution des réseaux, 
dynamique des réseaux, 
propriétés macroscopiques, 
réseaux de télécommunications
}

\keywords{
Social networks, 
complex networks, 
network evolution, 
network dynamics, 
macroscopic properties, 
telecommunication networks
}

\maketitlepage

\section{Introduction}
\input{section/introduction.tex}


\section{Evolution des propriétés globales} \label{secGlobal}
\input{section/global.tex}


\section{Comportements locaux de formation de liens} \label{secLocal}
\input{section/local.tex}


\section{Conclusion} \label{secConclusion}
\input{section/conclusion.tex}


\bibliographystyle{plain}
\bibliography{biblio}
\end{document}

%% file: section/abstractFr.tex
Les modèles classiques de génération de réseaux tentent de reproduire des propriétés structurelles globales (distribution des degrés, distance moyenne, coefficient de clustering, communautés, etc.) observées sur les réseaux du monde réel, par le biais de mécanismes de formation de liens synthétiques tels que la fermeture triadique ou l'attachement préférentiel.
Dans ce travail, nous étudions l'évolution d'un très grand réseau de communications en téléphonie mobile et nous analysons le processus de formation des liens.
Une première étude permet d'observer que plusieurs mécanismes sont à l'origine des propriétés globales observées sur ce réseau et qu'ils ne correspondent pas tous au schéma classique.
Dans une seconde étude, nous caractérisons plus finement le processus de formation des liens en recherchant des corrélations entre la probabilité de créer un nouveau lien et certaines propriétés individuelles telles que le degré, le coefficient de clustering et l'âge des noeuds.

%% file: section/abstractEn.tex
Traditional network generation models attempt to replicate global structural properties (degree distribution, average distance, clustering coefficient, communities, etc.) through synthetic link formation mechanisms such as triadic closure or preferential attachment.
In this work, we study the evolution of a very big communication network coming from mobile telephony and we analyse the link formation process.
A first study conducted on the standard mechanisms allows observing that several mechanisms are responsible for the properties observed in this network.
In a second study, we characterize more precisely the link formation process by searching for correlations between the probability of creating a new link and some individual properties such as the degree, the clustering coefficient and the age of the nodes.

%% file: section/introduction.tex
Les réseaux du monde réel ont souvent des structures topologiques non-triviales qui émergent des interactions entre les individus, on parle ainsi de \textit{réseaux complexes}~\cite{Boccaletti2006}.
Ces dernières années, l'étude de l'évolution de ces réseaux a été un domaine de recherche très actif~\cite{Barabasi1999, Dorogovtsev2002, Boguna2003, Stattner2012-CHB}. 
En effet, de nombreux modèles de génération ont été proposés pour reproduire les propriétés topologiques particulières observées sur les réseaux du monde réel (scale-free, petit-monde, structure communautaire forte, etc.).

La plupart de ces modèles sont basés sur l'hypothèse que ces propriétés globales sont le résultat de mécanismes d'évolution élémentaires gouvernant les changements structurels survenant au sein de l'ensemble du réseau~\cite{Barabasi2002, Kumpula2007, Toivonen2009}.
Par exemple, des réseaux possédant des coefficients de clustering élevés sont reproduits en appliquant le mécanisme dit de \textit{fermeture triadique}~\cite{Opsahl2011}, c'est-à-dire privilégiant la connexion entre des noeuds possédant des voisins communs.
De même, partant d'un réseau régulier, Watts et Strogatz ont proposé un modèle~\cite{Watts1998} basé sur une \textit{formation aléatoire} des liens pour générer des réseaux de type petit-monde.
Les réseaux scale-free peuvent, eux, être reproduits en utilisant le modèle de Barabasi et Albert~\cite{Barabasi1999} qui introduit la notion \textit{d'attachement préférentiel}, c'est-à-dire que le processus de création de liens privilégie la formation avec des individus déjà fortement connectés.

Toutefois, bien que ces mécanismes permettent d'obtenir des réseaux possédant des propriétés structurelles globales identiques à celles observées sur les réseaux complexes, ils ne reflètent pas les changements microscopiques réels intervenant sur les réseaux du monde. Il est en effet peu probable que tous les individus d'un même réseau créent des liens selon le même principe.
C'est la raison pour laquelle d'autres travaux se sont intéressés à des modèles de génération faisant intervenir plusieurs mécanismes d'évolution.
Toivonen et al.~\cite{Toivonen2009} ont par exemple comparé plusieurs modèles combinant fermeture triadique, attachement préférentiel et formation aléatoire, et ont montré qu'ils étaient en mesure de reproduire des réseaux de natures diverses.

Ainsi, les études récentes menées sur l'évolution de réseaux réels ont montré que les changements structurels étaient dirigés par une combinaison de mécanismes aussi bien endogènes, qu'exogènes au réseau, liés notamment aux comportements individuels des noeuds~\cite{Kossinets2006, Stattner2012-ANT}.
Par exemple, en étudiant plusieurs réseaux issus de sites d'échanges et de partages (Flickr, Delicious, Answers et LinkedIn), Leskovec et al.~\cite{Leskovec2008} ont pu observer que les mécanismes traditionnellement utilisés pour la génération de réseaux complexes, tels que la fermeture triadique ou l'attachement préférentiel, s'observaient finalement très peu.
Selon eux, "\textit{la nature intrinsèquement non-local de l'attachement préférentiel est fondamentalement incapable de capturer les caractéristiques importantes de ces réseaux}".

Dans ce travail, nous étudions l'évolution d'un très grand réseau de télécommunications issus de la téléphonie mobile et nous analysons le processus de formation de liens.
Notre objectif est de comprendre et de caractériser, qualitativement et quantitativement, les dynamiques impliquées dans la formation. 
\\
Ainsi, une première étude menée sur l'évolution globale du réseau permet d'effectuer deux observations intéressantes:
(i)~comme de nombreux réseaux du monde réel, ce réseau conserve la propriété "scale-free" durant son évolution,
(ii)~les mécanismes de formations élémentaires, couramment utilisés pour simuler l'évolution des réseaux complexes ne sont impliqués que dans de faibles proportions dans l'évolution de ce réseau.
Dans une deuxième étude, nous cherchons à caractériser plus finement le processus de formation en recherchant des corrélations entre la formation des nouveaux liens et certaines propriétés individuelles des noeuds telles que le degré, le coefficient de clustering et l'âge.

Cet article est organisé comme suit.
La Section~\ref{secGlobal} est dédiée à l'étude menée sur l'évolution globale du réseau.
Dans la Section~\ref{secLocal}, nous caractérisons plus précisément la formation de liens en recherchant les implications de certaines propriétés locales des noeuds.
Enfin, la Section~\ref{secConclusion} conclut cet article et présente nos travaux futurs.

%% file: section/global.tex
Dans une première étude, nous nous sommes intéressés à l'évolution du réseau et nous avons cherché à comprendre si la formation des liens au sein de ce réseau suivait les schémas traditionnellement mis en place dans les modèles de génération.

Dans la Section~\ref{subDataset} nous présentons le jeu de données utilisé et nous montrons comment évolue ses principales propriétés structurelles au cours du temps.
Dans la Section~\ref{subMotifsTraditionnels}, nous nous intéressons plus précisément à l'apparition des liens du réseau, et nous cherchons à déterminer si ces liens correspondent aux motifs traditionnellement utilisés pour la génération de réseaux complexes.

\subsection{Jeu de données et évolution} \label{subDataset}
Le jeu de données, fourni par un opérateur de téléphonie mobile local, consiste en un ensemble de fichiers représentant les communications sur 7~mois (Juin à Décembre 2009).
Chaque fichier regroupe l'ensemble des communications effectuées sur une journée.
Enfin dans un fichier, chaque ligne détaille une communication, en précisant notamment le numéro de \textit{l'appelant} et de \textit{l'appelé}, \textit{le type d'échange} (Appel, SMS ou Fax), \textit{la durée de la communication}, etc.

La première étape de notre travail a consisté à traiter ces données pour en extraire le réseau de communications sous-jacent entre les abonnés (limité aux appels et SMS).
Dans ce réseau, \textit{un noeud} représente un abonné, et \textit{un lien} une communication.
Ainsi, nous pouvons étudier comment évolue ce réseau jour après jour en analysant les fichiers séquentiellement.
En raison de la quantité de données à traiter, et du temps de calculs associé (ce dernier aspect est détaillé dans la section suivante), nous nous sommes limités pour ce travail préliminaire à l'étude de l'évolution du réseau sur 2~semaines.
L'approche que nous avons adoptée était la suivante: le réseau était d'abord construit sur 3 jours, puis analysé sur 15, soit 18 fichiers au total.

\begin{figure}[!h]
	\centering
	\subfigure[]{
		\includegraphics[scale=0.2]{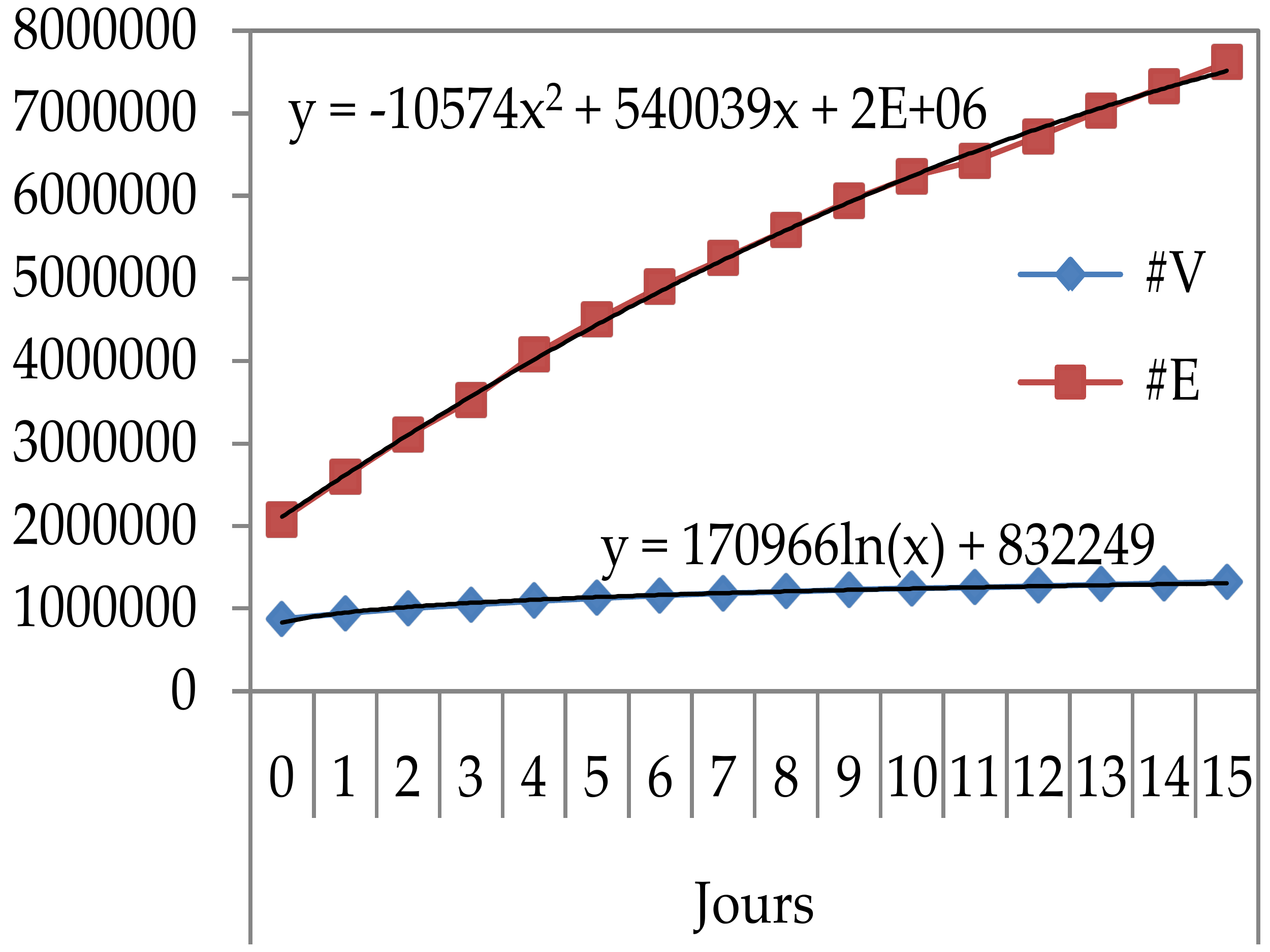}
	}
	\subfigure[]{
		\includegraphics[scale=0.2]{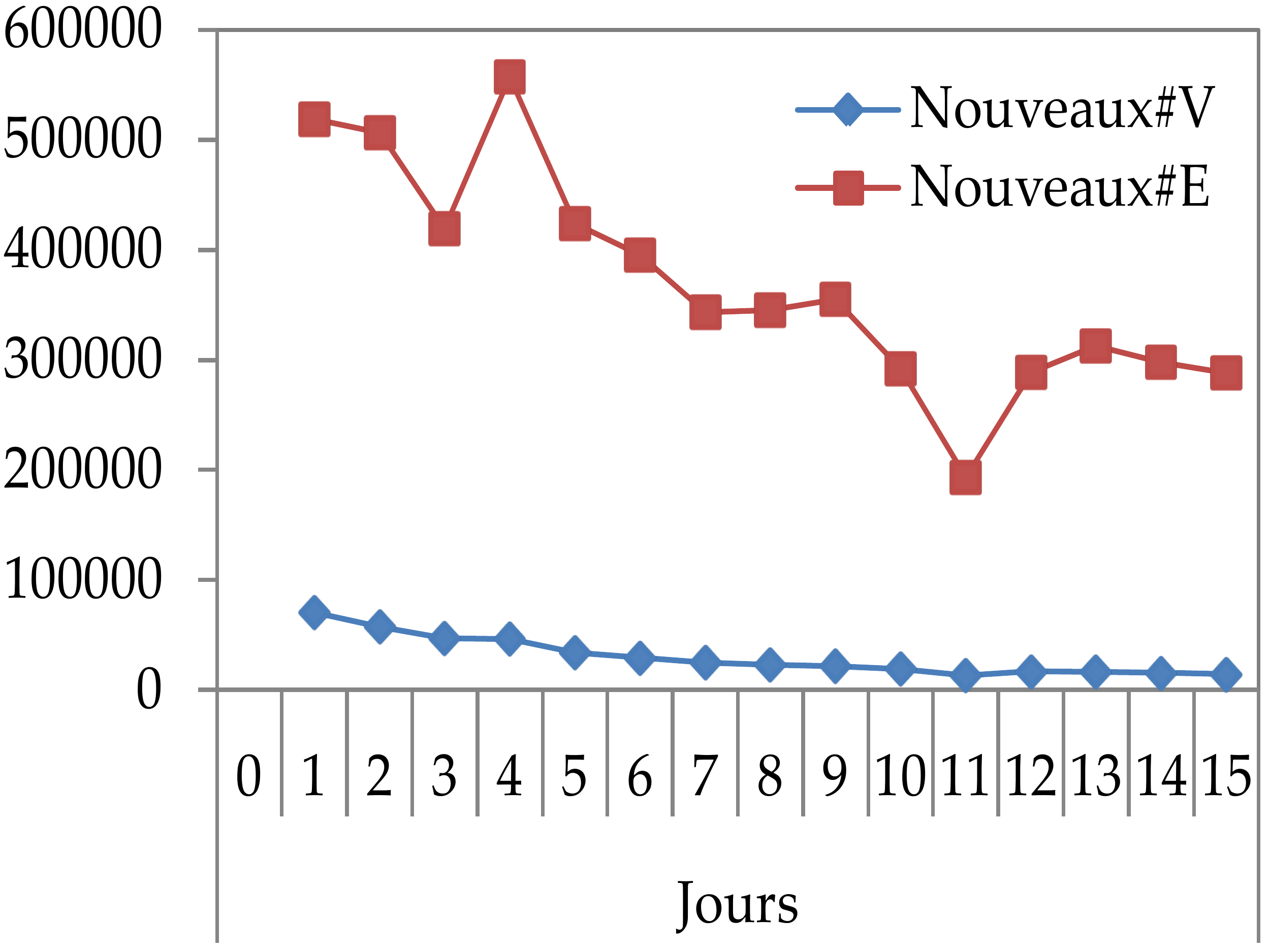}
	}
	\subfigure[]{
		\includegraphics[scale=0.2]{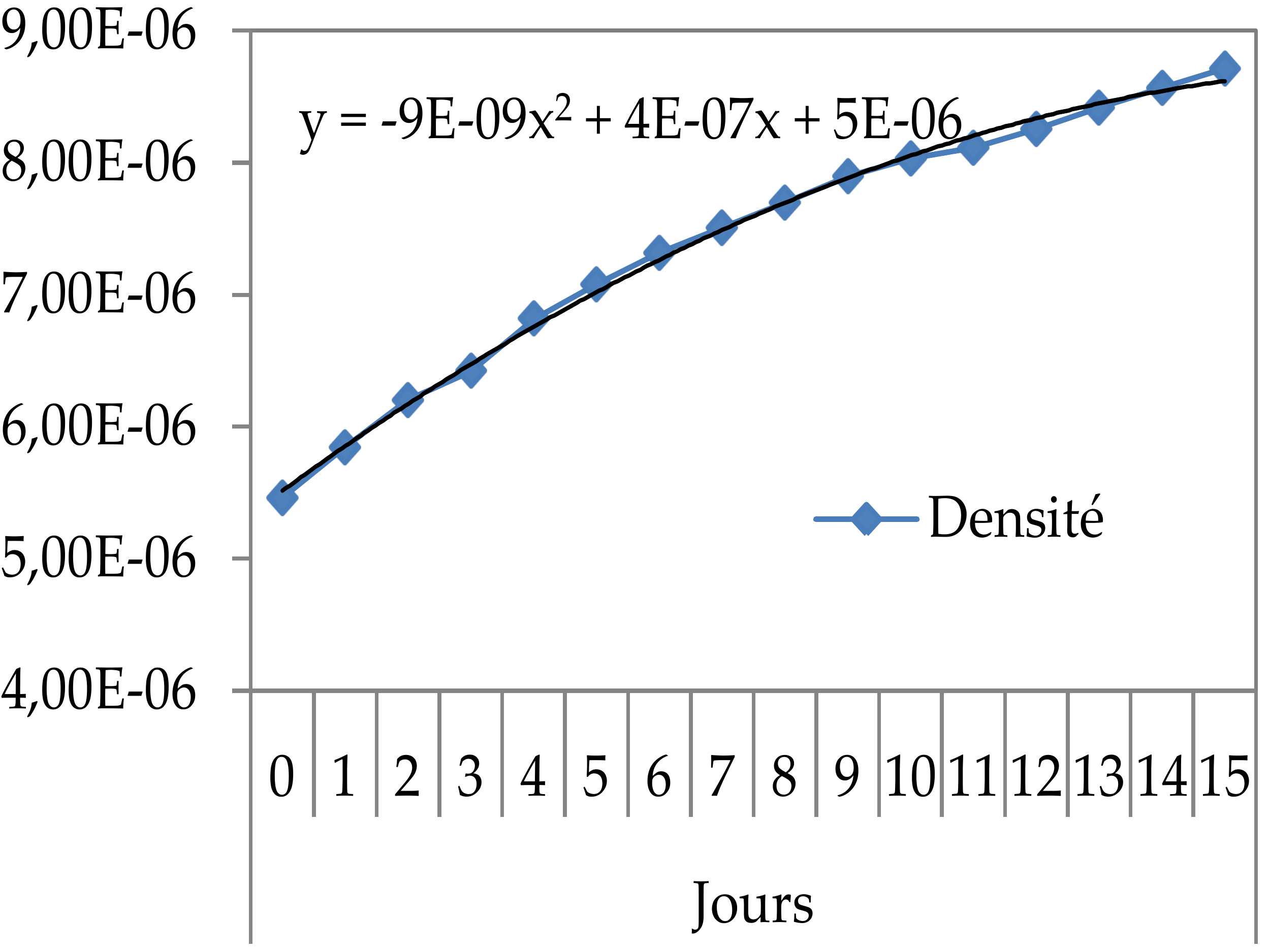}
	}
	\subfigure[]{
		\includegraphics[scale=0.2]{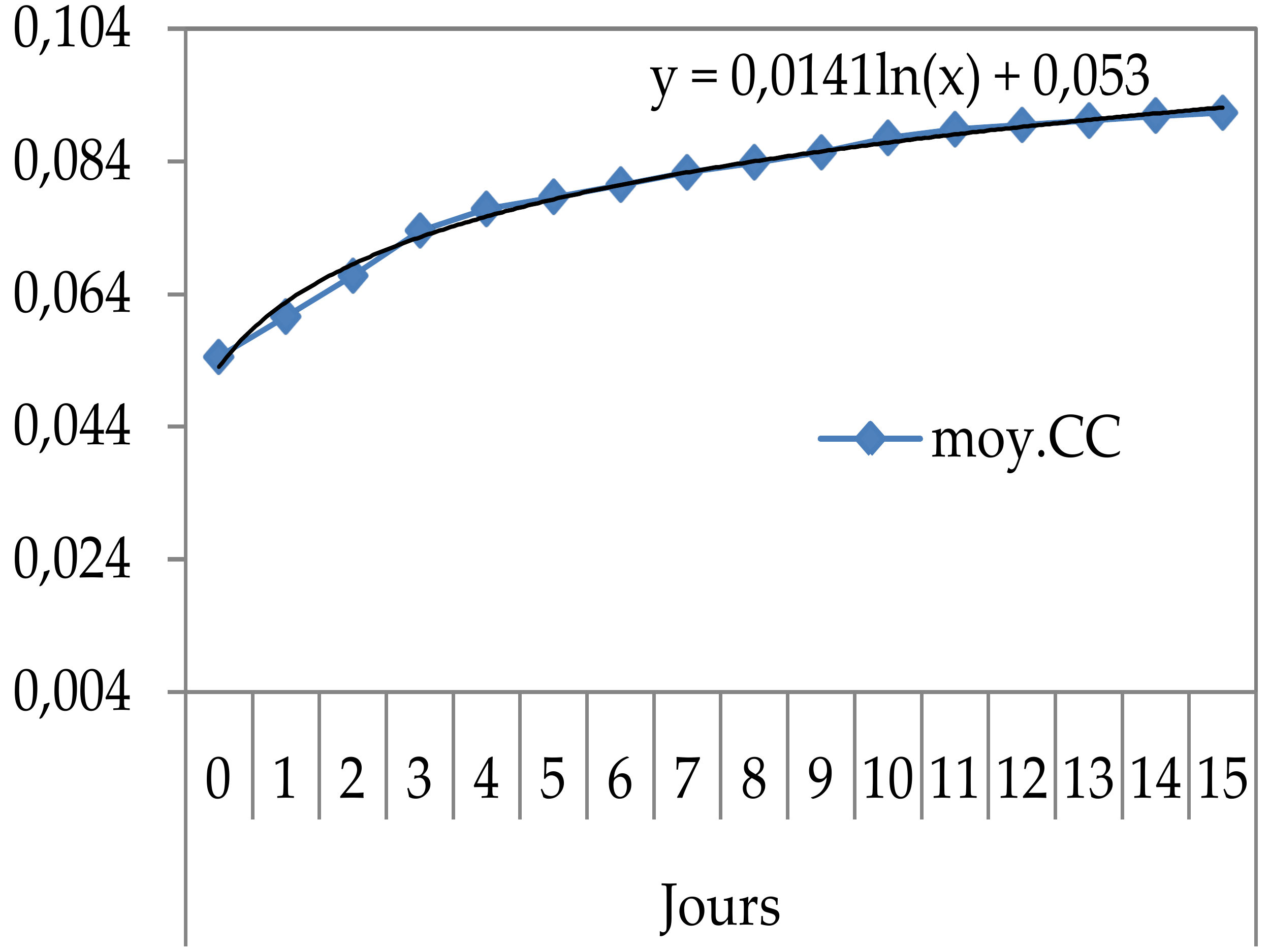}
	}
	\caption
	{Évolution des principales propriétés du réseau sur 15 jours}
	{(a)~nombre de noeuds et de liens
	(b)~nombre de nouveaux noeuds et de nouveaux liens
	(c)~densité et 
	(d)~coefficient de clustering moyen.}
	\label{figProprietes_1}
\end{figure}

Dans un premier temps, nous nous sommes intéressés à l'évolution des principales propriétés du réseau:
(a)~nombre de noeuds $\#V$ et de liens $\#E$, 
(b)~nombre de nouveaux noeuds et de nouveaux liens distincts, 
(c)~densité et 
(d)~coefficient de clustering moyen.
La Figure~\ref{figProprietes_1} montre comment évolue ces caractéristiques sur 15~jours.

Comme attendu, le nombre de noeuds et de liens connaissent tous deux des évolutions croissantes sur la période d'étude (cf Figure~\ref{figProprietes_1}(a)).
Cela s'explique par le fait qu'au début de l'analyse, le réseau considéré, bien que résultat d'un cumul sur 3 jours, ne représente pas toutes les communications existantes et est donc peu fourni.
Une grande quantité de noeuds et de liens encore non-identifiés est ainsi insérée au réseau.
On observe tout de même que cette tendance tend a diminuer pour le nombre de noeuds, puisqu'on constate une phase stabilisation de l'évolution qui peut être approchée par une fonction logarithmique.
Ces résultats laissent ainsi penser que deux semaines suffisent pour que la grande majorité des abonnés utilisant leur forfait soit impliquée dans au moins une communication, ce qui permet de les ajouter au réseau.
\\
Les résultats obtenus pour le nombre de nouveaux noeuds et de liens confirment nos observations précédentes (cf Figure~\ref{figProprietes_1}(b)).
Bien que le réseau ne fasse que croître, les liens n'étant jamais supprimés, on observe que le nombre de nouveaux noeuds et de nouveaux liens, eux, décroissent.
La stabilisation du nombre de noeuds $\#V$ observée précédemment se traduit ici par une décroissance très forte du nombre de nouveaux noeuds qui semble tendre vers $0$.
Ce qui confirme qu'une grande partie des abonnés a été insérée au réseau.
\\
Enfin, la densité et le coefficient de clustering moyen (cf Figures~\ref{figProprietes_1}(a) et (d)) sont tous les deux croissants sur la période.
On observe d'ailleurs que le coefficient de clustering tend à se stabiliser, ce qui suggère que l'ajout de nouveaux liens crée peu de "triangles" dans le réseau et donc ne tend pas à renforcer l'effet communautaire dans ce réseau.

\begin{figure}[!h]
	\centering
	\subfigure[]{
		\includegraphics[scale=0.2]{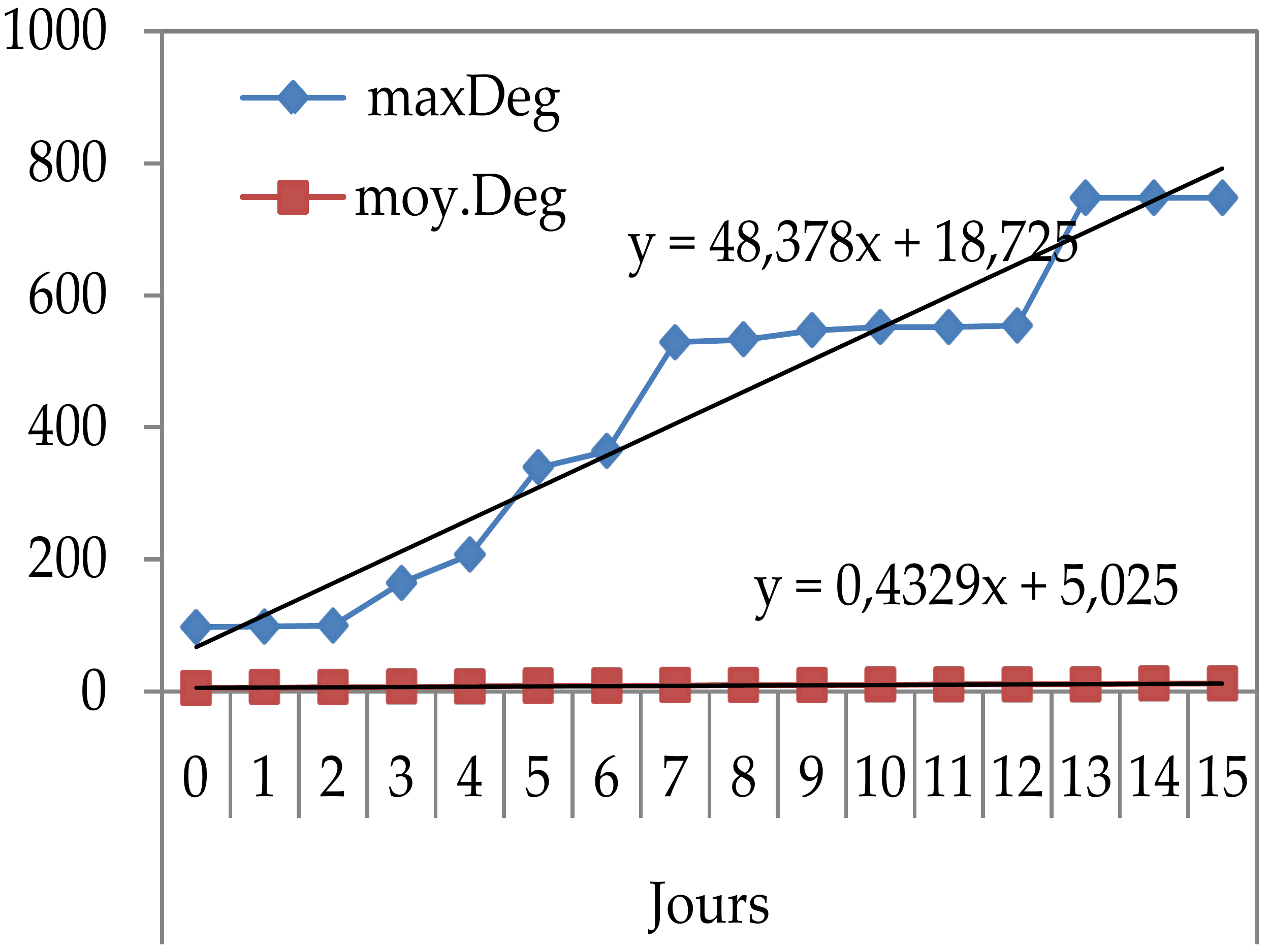}
	}
	\subfigure[]{
		\includegraphics[width=6cm,height=4cm]{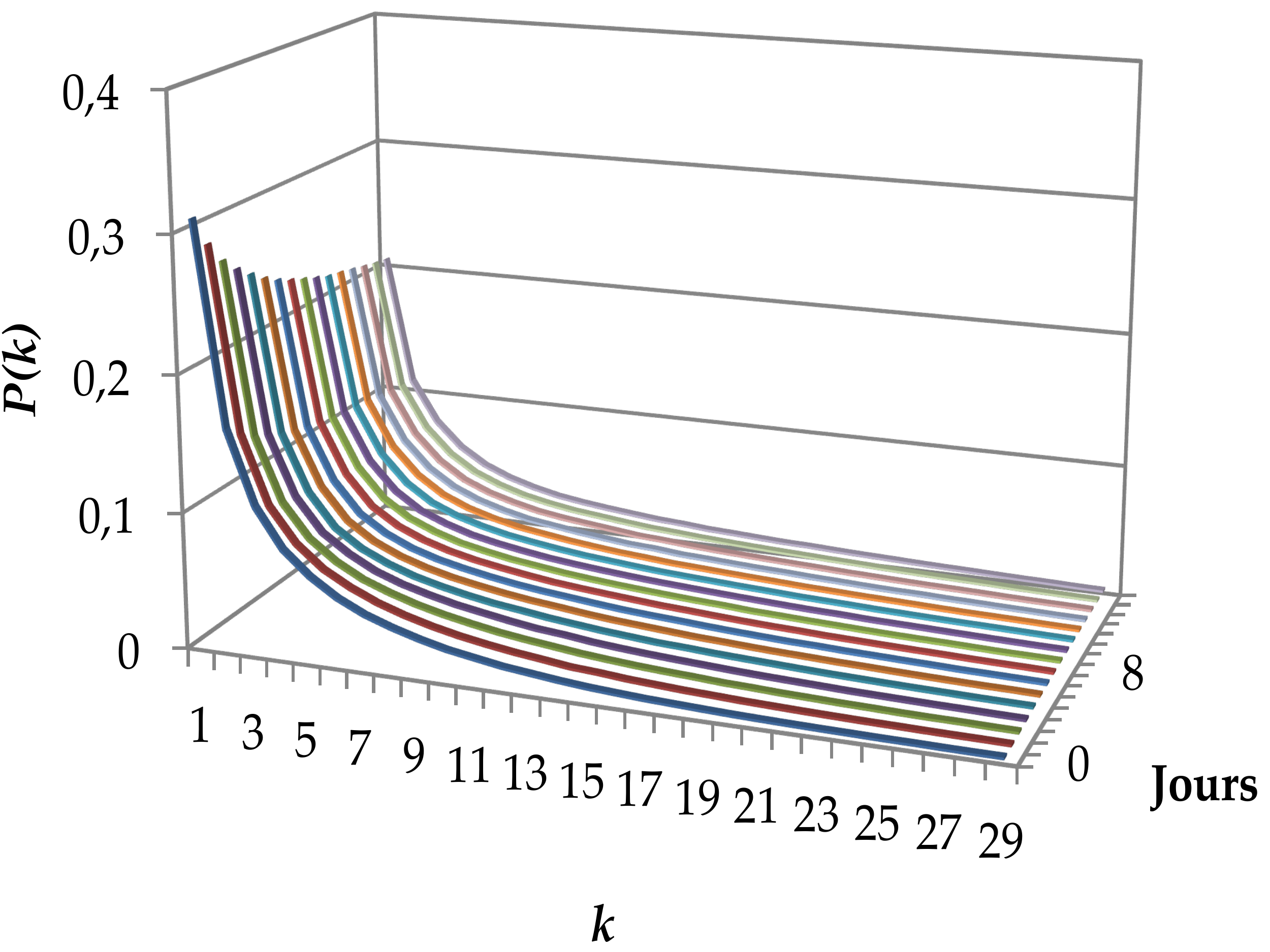}
	}
	\caption{Évolution des degrés du réseau sur 15 jours: (a)~degré moyen et degré max et (b)~distribution des degrés}
	\label{figProprietes_2}
\end{figure}

Enfin, pour compléter cette étude sur l'évolution des propriétés, nous nous sommes intéressés à l'évolution de plusieurs mesures relatives au degré des noeuds, de façon à vérifier si ce réseau possédait une structure connue.
La Figure~\ref{figProprietes_2} montre comment évolue
(a)~la degré moyen et le degré max et 
(b)~la distribution des degrés. 

Il est intéressant d'observer l'écart considérable qui existe entre le degré moyen et le degré max (cf Figure~\ref{figProprietes_2}(a)). Cela laisse supposer qu'il existe un très petit nombre d'individus extrêmement connectés dans ce réseau.
Les résultats obtenus pour l'évolution de la distribution des degrés  (cf Figure~\ref{figProprietes_2}(b)) confirme cette hypothèse et permettent d'observer que le réseau conserve une structure de type \textit{scale-free} tout au long de son évolution.

\subsection{Motifs d'évolution traditionnels} \label{subMotifsTraditionnels}
L'étude précédente nous a permis d'observer que le réseau de télécommunications étudié conservait une structure scale-free lors son évolution.
Or, nous savons que cette propriété est souvent expliquée par le mécanisme dit "\textit{d'attachement préférentiel}".

Ainsi, soit $G=(V,E)$ un réseau, avec $V$ l'ensemble des noeuds et $E \subseteq V \times V$ l'ensemble des liens.
On note $k_{v_{i}}$ le degré du noeud $v_{i}$.
Selon le modèle proposé par Barabasi et Albert, la probabilité $p_{v_{i}}$ de créer un lien avec un noeud $v_{i}$ s'obtient par:
\begin{equation}
	p_{v_{i}} = \frac{k_{v_{i}}}{\sum_{j} k_{v_{j}}}
\end{equation}
où $\sum_{j} k_{v_{j}}$ correspond à la somme des degrés des autres noeuds du réseau. Ainsi, la probabilité qu'a le noeud $v_{i}$ d'être impliqué dans la formation d'un nouveau lien croit proportionnellement avec son degré.

\begin{figure}[!h]
	\centering
	\includegraphics[scale=0.2]{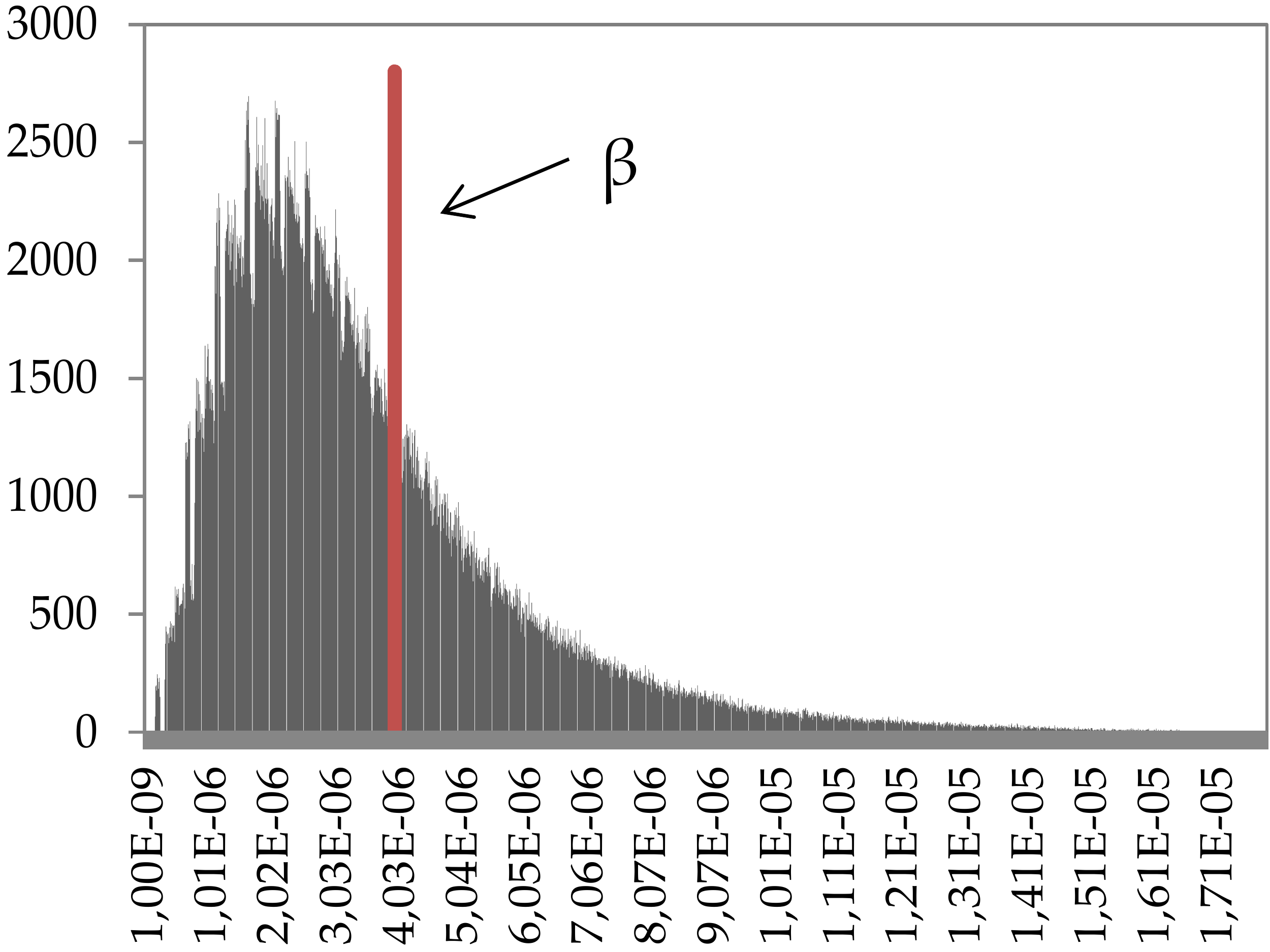}
	\caption{Distribution des $p_{v_{i}}$ lors de la création de liens sur 15 jours}
	\label{figPatterns_1}
\end{figure}

Dans un second temps, nous nous sommes intéressés au processus de formation des liens au sein de ce réseau de communications entre abonnées, et plus précisément, nous avons cherché à comprendre si la structure scale-free observée s'expliquait uniquement par le mécanisme d'attachement préférentiel.
Ainsi, pour chaque lien créé durant les 15 jours d'étude, nous avons calculé la probabilité $p_{v_{i}}$ qu'il aurait eu dans le modèle de Barabasi et Albert.
La Figure~\ref{figPatterns_1} montre la distribution des $p_{v_{i}}$ après 15 jours.

Il est particulièrement intéressant d'observer que le nombre de liens à l'origine des $p_{v_{i}}$ élevés, c'est-à-dire assimilables à de l'attachement préférentiel, est très faible. 
La grande majorité des créations est en effet associée à des $p_{v_{i}}$ relativement faibles, ce qui suggère que le processus d'attachement préférentiel dans ce réseau s'avère finalement être assez marginal.

Ainsi, malgré la propriété scale-free que conserve ce réseau durant son évolution, l'attachement préférentiel ne semble pas être l'unique processus à l'origine de la formation des liens.
Bien qu'on puisse raisonnablement penser que peu de liens de type attachement préférentiel suffisent à maintenir cette structure scale-free, la question centrale concerne les processus impliqués dans la formation des autres liens.
\\
Posons $\beta$ un \textit{seuil de probabilité} à partir duquel on considère que le lien est de type attachement préférentiel.
Pour mieux comprendre comment évolue ce réseau, nous avons tenté de classifier chaque lien selon les mécanismes de formation traditionnellement utilisés dans la littérature:
\begin{itemize}
	\item \textit{Attachement préférentiel} (PA), qui fait référence au mécanisme de formation qui privilégie la création d'un lien avec un individu déjà fortement connecté.
	\item \textit{Fermeture triadique} (TC), qui correspond au mécanisme à travers lequel un noeud crée un nouveau lien avec les amis de ses amis.
	\item \textit{Aléatoire} (R): concerne tous les liens qui ne peuvent pas être classés dans une des deux classes précédentes et qu'on suppose aléatoires.
\end{itemize}
L'Algorithme~\ref{algoIdentification} détaille le principe d'identification des liens. 
Nous pouvons notamment observer que le type associé à un lien n'est pas exclusif, puisqu'un lien peut être à la fois identifié comme étant de type attachement préférentiel et de type fermeture triadique.

\begin{algorithm}[!t]
	\small
	\caption{Identification du type de formation de liens}
	\label{algoIdentification}
	
	\begin{algorithmic}[1]
		\algsetup{linenosize=\small, linenodelimiter=.}
		
		\REQUIRE $G=(V,E)$: \textbf{Réseau}, $\beta$: \textbf{seuil} $\in [0..1]$ 
		
		\STATE $PA$: Entier $\leftarrow 0$
		\STATE $TC$: Entier $\leftarrow 0$
		\STATE $R$: Entier $\leftarrow 0$ 
		\FORALL{lien $e=(v_{i},v_{j})$ lu dans les fichiers}
			\IF{$e \notin E$}
				\STATE $r1$: Booléen $\leftarrow$ faux
				\STATE $r2$: Booléen $\leftarrow$ faux
				\IF{$p_{v_{j}} \geq \beta$}
					\STATE $r1$ $\leftarrow$ vrai
					\STATE $PA \leftarrow PA+1$
				\ENDIF
				\IF{$v_{i}$ et $v_{j}$ ont au moins un voisin commun}
					\STATE $r2$ $\leftarrow$ vrai
					\STATE $TC \leftarrow TC+1$
				\ENDIF
				\IF{$r1 =$ faux ET $r2 =$ faux}
					\STATE $R \leftarrow R+1$
				\ENDIF
				\STATE Ajouter $e$ à $E$
			\ENDIF
		\ENDFOR
		\end{algorithmic}
\end{algorithm}

\begin{figure}[!h]
	\centering
	\subfigure[]{
		\includegraphics[scale=0.2]{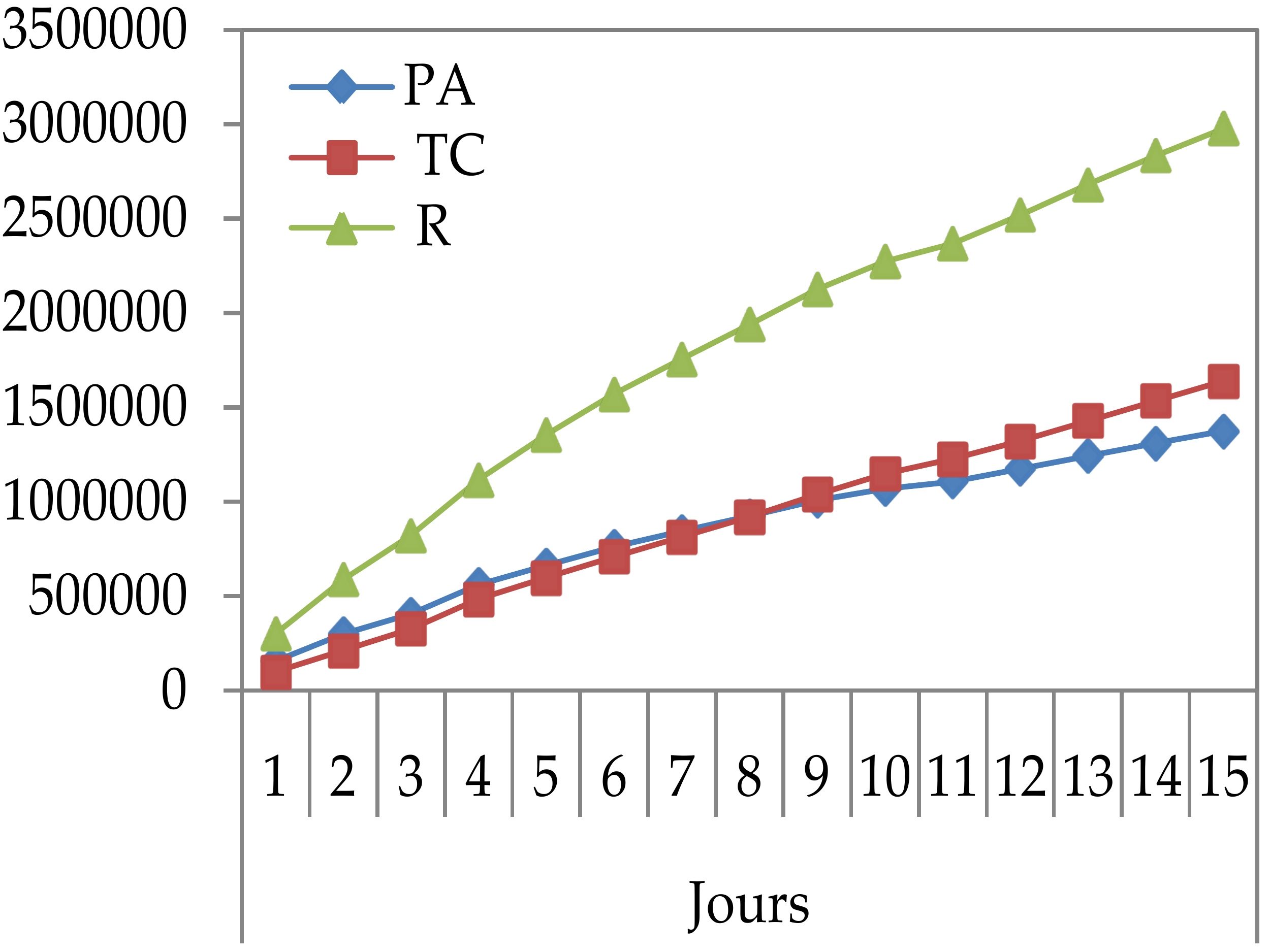}
	}
	\subfigure[]{
		\includegraphics[scale=0.2]{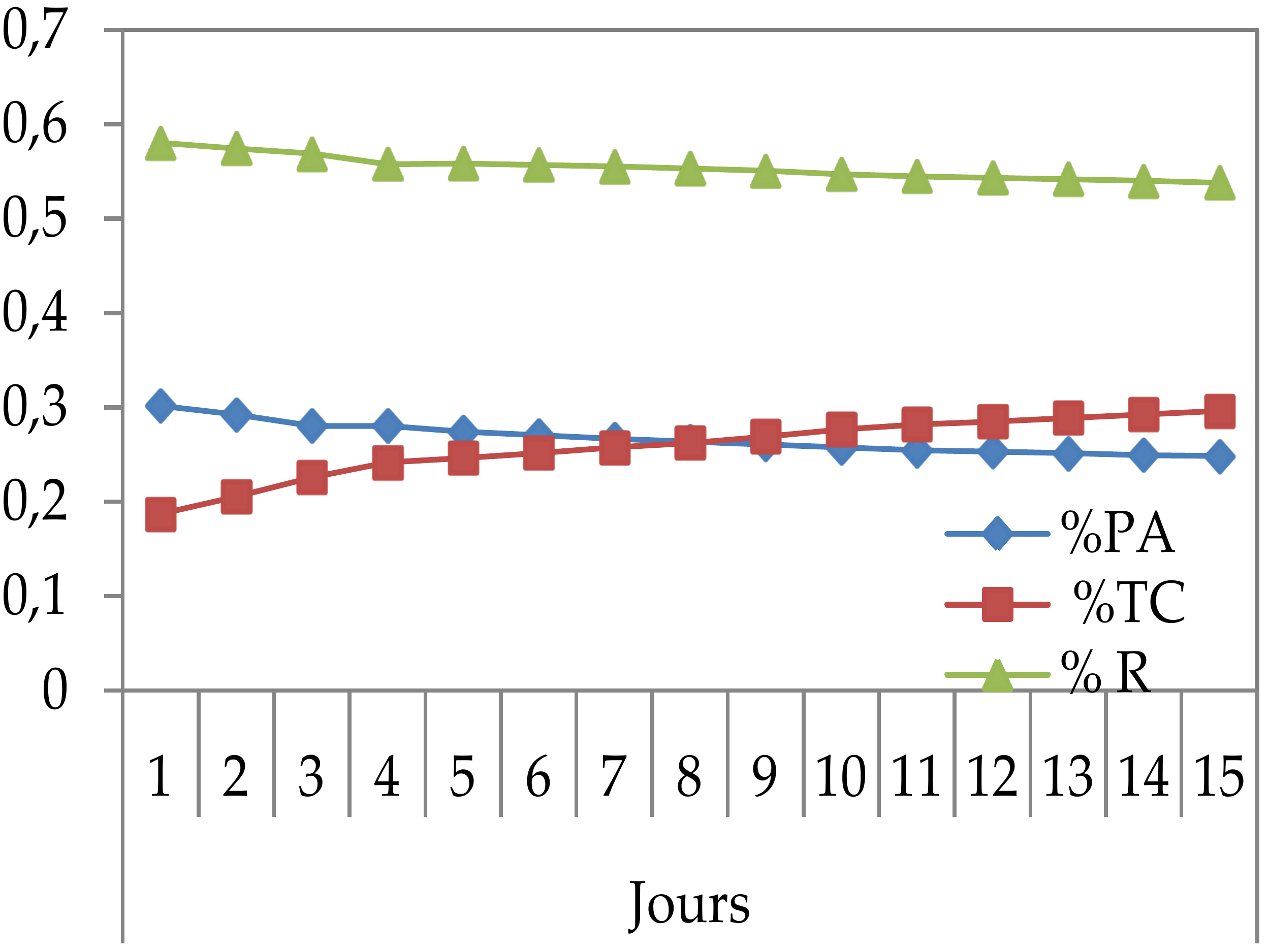}
	}
	\caption{Évolution du nombre de liens de type PA, TC et R (a)~en quantité et (b)~en proportion}
	\label{figPatterns_2}
\end{figure}

En fixant $\beta = 4,00E$-$06$ (représenté sur la Figure~\ref{figPatterns_1}), la Figure~\ref{figPatterns_2} montre l'évolution du nombre de liens de type PA, TC et R sur les 15~jours (a)~en quantité et (b)~en proportion.

Comme observé précédemment, on constate que le nombre de liens se formant suivant le mécanisme PA reste finalement assez faible dans le réseau. Seul $20\%$ à $30\%$ des liens sont en effet identifiés comme étant de l'attachement préférentiel.
Nous pouvons également observer que les liens de type TC sont présents dans des proportions similaires.
Ce qui nous permet d'affirmer que dans l'évolution de ce réseau, il y a globalement autant de liens se formant selon l'attachement préférentiel que de liens se formant selon la fermeture triadique, et ce, quelque soit le moment de l'observation.
On peut également noter que ces proportions restent relativement constantes dans le temps.
Ces résultats viennent ainsi confirmer deux observations également faites par Leskovec et al.~\cite{Leskovec2008}, qui montraient qu'environ $30\%$ des nouveaux liens formés dans les réseaux qu'ils étudiaient étaient de type fermeture triadique.

L'observation la plus intéressante concerne sans doute le fort pourcentage de liens non-identifiés.
En effet, $60\%$ des nouveaux liens ne sont identifiés ni comme étant de l'attachement préférentiel, ni comme étant de la fermeture triadique, ce qui nous amène à nous interroger sur la nature de ces liens, en nous intéressant notamment aux implications de certaines propriétés locales dans le processus de formation.

%% file: section/local.tex
Pour aller plus loin dans l'étude des nouveaux liens apparaissant dans le réseau, nous nous sommes, dans cette seconde étape, intéressés aux corrélations entre les propriétés individuelles des noeuds et le processus de formation des liens.
\\
Nous avons ainsi caractérisé les nouveaux liens, selon trois grandes caractéristiques des noeuds: 
(1)~le degré,
(2)~le coefficient de clustering et
(3)~l'âge (calculé en fonction du jour d'apparition du noeud dans le réseau).
Pour chacune d'elles, nous définissons trois types de formation des liens, comme décrit ci-dessous:

\begin{enumerate}
	\item \textbf{Le degré:}
	\begin{itemize}
		\item $DHH$ correspond à un lien formé entre deux noeuds hautement connectés au regard du degré moyen, c'est-à-dire un lien $e=(v_{i},v_{j})$ est de type DHH si $k_{v_{i}} \geq Moy.Deg$ et $k_{v_{j}} \geq Moy.Deg$.
		\item $DLL$ défini un lien formé entre deux noeuds faiblement connectés au regard du degré moyen, c'est-à-dire un lien $e=(v_{i},v_{j})$ est de type DLL si $k_{v_{i}} < Moy.Deg$ et $k_{v_{j}} < Moy.Deg$.
		\item $DHL$ correspond à un lien formé entre des noeuds mixtes au regard du degré moyen, c'est-à-dire si ($k_{v_{i}} < Moy.Deg$ et $k_{v_{j}} \geq Moy.Deg$) ou ($k_{v_{i}} \geq Moy.Deg$ et $k_{v_{j}} < Moy.Deg$).
	\end{itemize}
	
	\item \textbf{Le coefficient de clustering:}
	\begin{itemize}
		\item $CHH$ correspond à un lien formé entre deux noeuds dont les coefficients de clustering sont supérieurs à la moyenne. Plus précisément, soit $cc(v_{i})$ le coefficient de clustering du noeud $v_{i}$, un lien $e=(v_{i},v_{j})$ est de type $CHH$ si $cc(v_{i}) \geq Moy.CC$ et $cc(v_{j}) \geq Moy.CC$.
		\item $CLL$ est un lien formé entre des noeuds possédant des coefficients de clustering inférieurs à la moyenne, i.e. $cc(v_{i}) < Moy.CC$ et $cc(v_{j}) < Moy.CC$.
		\item $CHL$ correspond à un lien formé entre des noeuds mixtes au regard du coefficient de clustering.
	\end{itemize}
	
	\item \textbf{L'age du noeud:}
	\begin{itemize}
		\item $AJJ$ défini un lien formé entre des noeuds identifiés comme "jeune" dans le réseau. Plus précisément, soient $t(v_{i})$ le jour où le noeud $v_{i}$ a été ajouté au réseau et $t(e)$ le jour où le lien a été ajouté, un lien $e=(v_{i},v_{j})$ est de type $AJJ$ si $t(v_{i}) \geq t(e) - 3~jours$ et $t(v_{j}) \geq t(e) - 3~jours$.
		\item $AOO$ défini un lien formé entre deux noeuds moins récents, c'est-à-dire
		un lien $e=(v_{i},v_{j})$ est de type $AOO$ si $t(v_{i}) < t(e) - 3~jours$ et $t(v_{j}) < t(e) - 3~jours$.
		\item $AJO$ correspond à un lien formé entre un jeune noeud et un noeud plus agé.
	\end{itemize}
\end{enumerate}

\begin{figure}[!h]
	\centering
	\subfigure[]{
		\includegraphics[width=11cm,height=4cm]{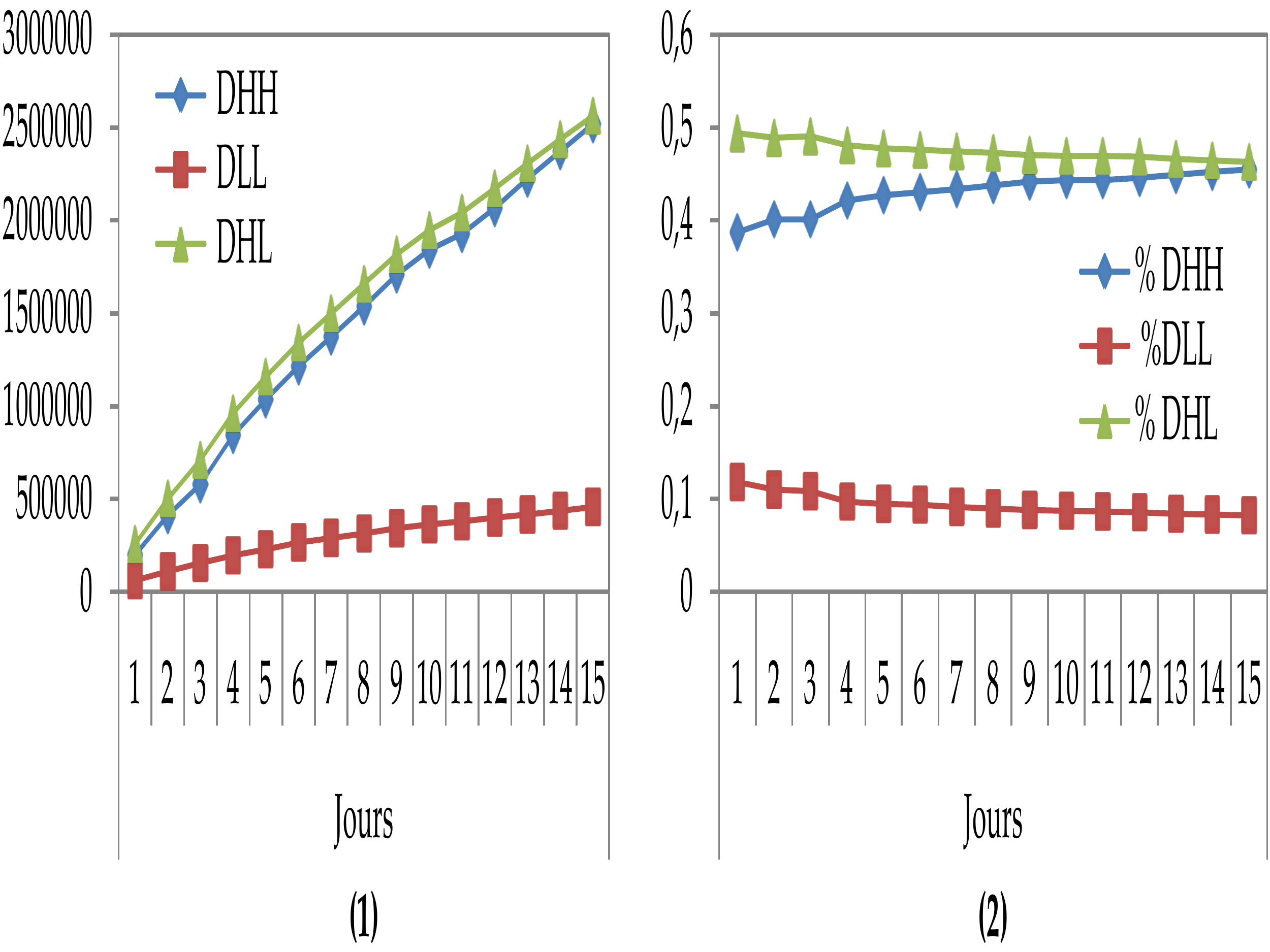}
	}
	\subfigure[]{
		\includegraphics[width=11cm,height=4cm]{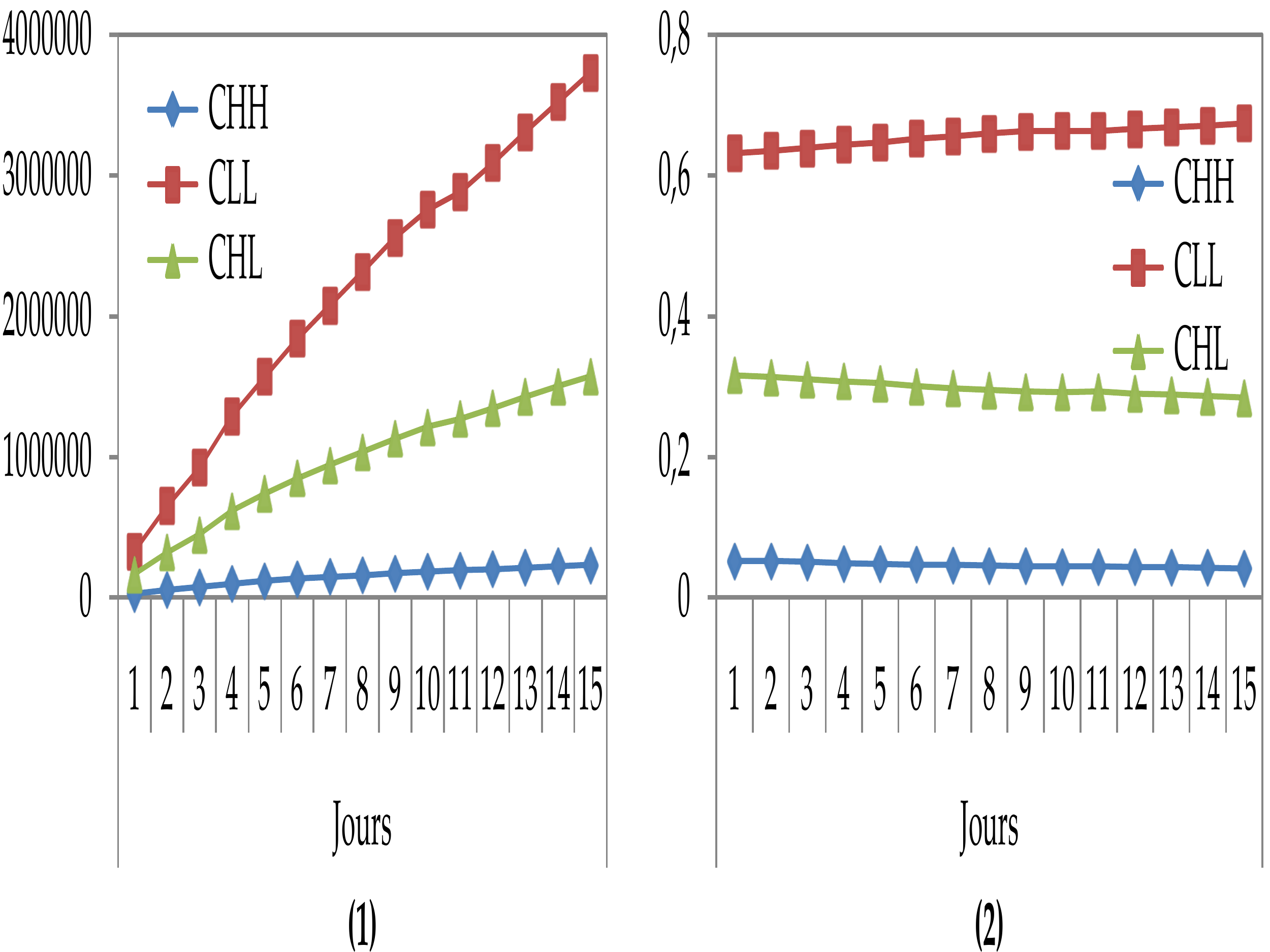}
	}
	\subfigure[]{
		\includegraphics[width=11cm,height=4cm]{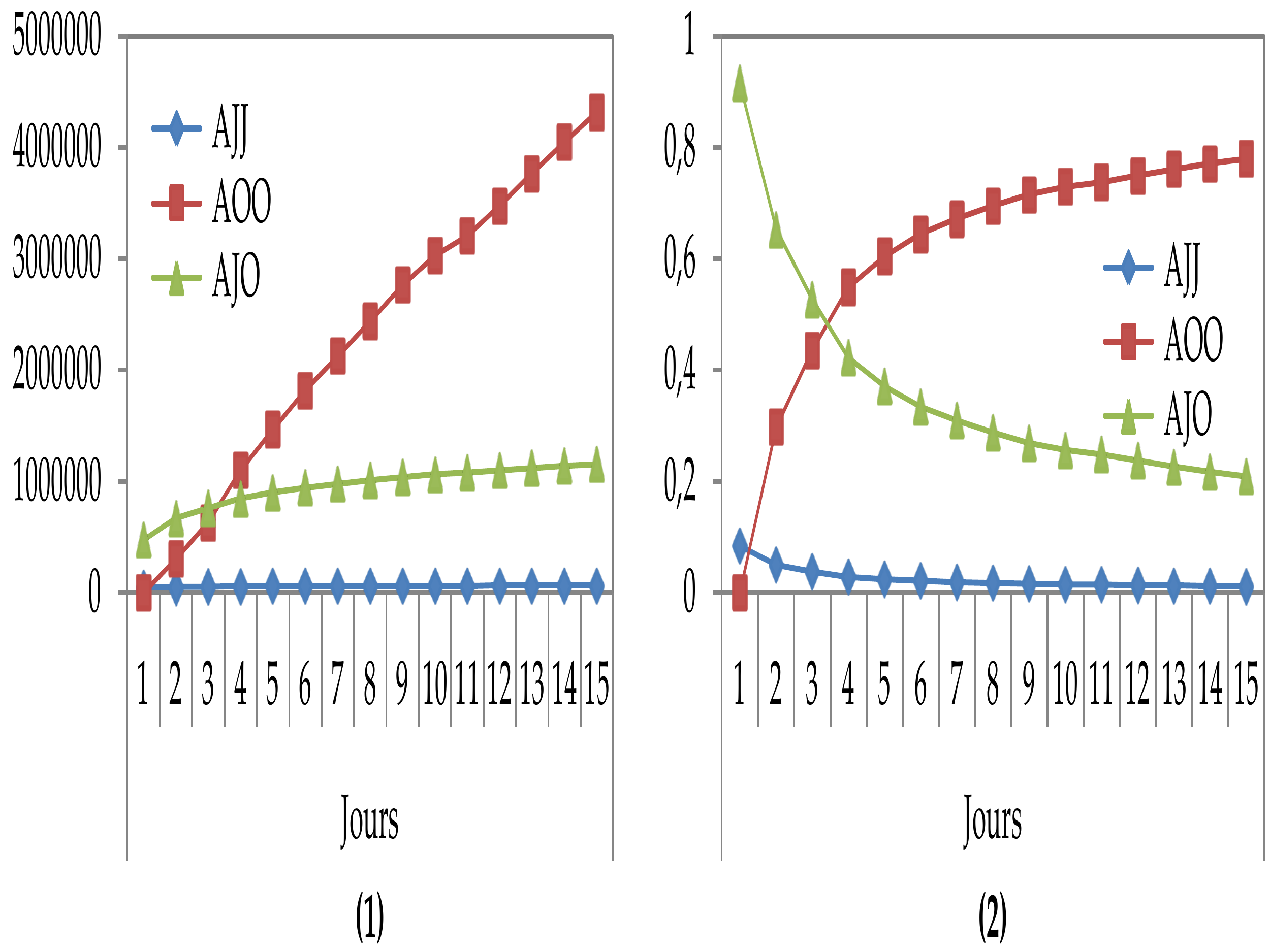}
	}
	\caption{Évolution des liens impliquant (a)~le degré (DHH, DLL et DHL), (b)~le coefficient de clustering (CHH, CLL et CHL) et (c)~l'age du noeud (AJJ, AOO, AJO): (1)~en quantitié et (2)~en proportion}
	\label{figLocal}
\end{figure}

La Figure~\ref{figLocal} montre comment évolue (1)~en quantitié et (2)~en proportion, les liens impliquant (a)~le degré (DHH, DLL et DHL), (b)~le coefficient de clustering (CHH, CLL et CHL) et (c)~l'age du noeud (AJJ, AOO, AJO).

Si l'expérience précédente a pu montrer que l'attachement préférentiel n'était pas le mécanisme dominant au sein de ce réseau, il est tout de même intéressant d'observer que la majeure partie des liens se crée entre des noeuds dont au moins un des deux possède un degré supérieur à la moyenne (cf Figures~\ref{figLocal}(a)).
En effet, $40\%$ à $45\%$ des liens se créent suivant DHH, alors que $45\%$ à $50\%$ des nouveaux liens s'avèrent être de type DHL.
Seuls $10\%$ des liens sont créés entre des noeuds possédant des degrés inférieurs à la moyenne.
Comme observé précédemment, ces proportions restent globalement stables durant l'évolution du réseau.
\\
On peut expliquer ces résultats par une forme "\textit{d'attachement préférentiel local}", à travers lequel les noeuds créent des liens avec les individus les plus connectés de leur entourage (ce qui explique le fort pourcentage pour DHH et DHL), mais qui ne sont pas nécessairement les plus connectés globalement d'où le faible pourcentage observé précédemment pour PA.

On peut également observer que la grande majorité des liens du réseau se crée entre des noeuds dont au moins un possède un coefficient de clustering inférieur à la moyenne (cf Figure~\ref{figLocal}(b)).
Plus précisément, $60\%$ des liens se créent entre deux noeuds possédant des coefficients de clustering inférieurs à la moyenne et $35\%$ se créent suivant le mécanisme $CHL$.
Seuls $5\%$ des liens se forment entre des individus possédant des coefficients de clustering supérieurs à la moyenne.
Ces tendances restent stables sur toute la durée de l'étude.

En ce qui concerne les liens impliquant l'age du noeud, les tendances sont intéressantes (cf Figures~\ref{figLocal}(c)).
On peut en effet observer que sur les trois premiers jours de l'étude, les liens de type AJO, c'est à dire impliquant des individus vieux et jeunes, sont très fréquents.
Par exemple, $90\%$ des nouvelles formations sont de type $AJO$ après le premier jour d'étude.
Ce résultat peut s'expliquer par le fait qu'au début de l'étude, de nombreux noeuds sont ajoutés au réseau, ce qui explique le taux de formation élevé avec des noeuds déjà présents.
Cependant, lorsque l'étude est prolongée, la tendance s'inverse et on observe que la plupart des liens se crée entre des noeuds identifiés comme vieux.
Par exemple, après le $7$e jour, $60\%$ à $70\%$ des nouveaux liens sont de type $AOO$, alors que $30\%$ à $20\%$ des liens se créent suivant le mécanisme $AJO$.
Seuls $10\%$ des nouveaux liens connectent des individus récemment ajoutés au réseau.

%% file: section/conclusion.tex
Dans ce travail, nous avons analysé l'évolution globale et locale d'un réseau de télécommunications.
L'étude menée est intéressante à plusieurs égards.
En effet, nous avons pu observer que lors de son évolution, le réseau conserve sa propriété scale-free mais n'est pas dirigé par un mécanisme de formation dominant tel que l'attachement préférentiel.
Nous avons ainsi pu mettre en évidence certaines corrélations entre les propriétés individuelles des noeuds et le processus de formation, ce qui nous amène à penser qu'un modèle de génération idéal serait un modèle capable de générer des réseaux possédant des propriétés structurelles particulières, tout en reproduisant localement les tendances de formation observées sur des réseaux réels.

Cette étude soulève également un certain nombre de questions sur la généralisation des résultats obtenus. 
Le réseau que nous avons utilisé est en effet un réseau possédant une sémantique bien particulière, puisqu'il s'agit de communications téléphoniques entre des individus.
On peut raisonnablement supposer que des études similaires menées sur des réseaux d'autres natures donneraient des résultats différents.
De telles études pourraient, à termes, permettre de proposer des modèles de génération ciblés, capables de reproduire l'évolution de réseaux de sémantique diverse: amitiés, collaboration, etc.

\begin{figure}[!t]
	\centering
	\includegraphics[width=6cm,height=4cm]{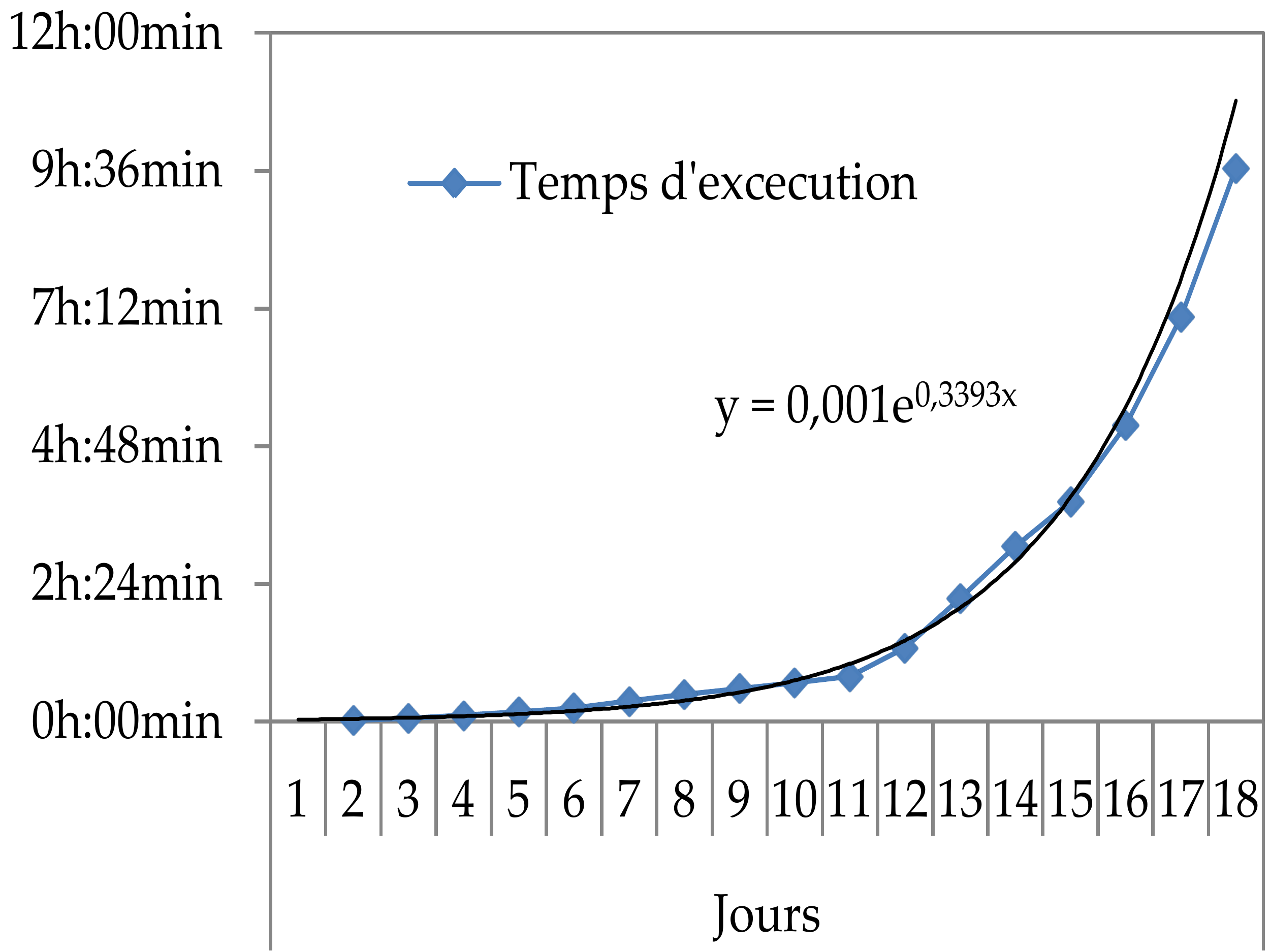}
	\caption{Évolution du temps de calcul en fonction du nombre de jours analysés}
	\label{figRuntime}
\end{figure}

Enfin, une des difficultés majeures de ce type d'étude concerne la durée d'observation et le passage à l'échelle.
Nous nous sommes en effet limité à l'analyse de l'évolution du réseau sur 15~jours en raison de l'énorme quantité de données à traiter et de la croissance exponentielle du temps de calcul avec la durée d'observation (voir Figure~\ref{figRuntime}; résultats obtenus avec Core~i7, 3,3GHz, 8Go de RAM, Java~7).
\\
Ainsi, un des défis majeurs à long terme consistera à mettre en place, pour l'étude de tels volumes de données, des solutions de collecte, de stockage et d'analyse efficientes de type "\textit{big data}", dans lesquelles la parallélisation des traitements est le facteur d'optimisation.

%% file: MARAMI2013.bbl
\begin{thebibliography}{}

\bibitem[BAR~99]{Barabasi1999}
\textsc{Barabasi A.}\andname{}\textsc{Albert R.}, \guilo{}Emergence of Scaling
  in Random Networks\guilf{},
\newblock \textit{Science}, \volumename\ 286(5439), 1999,  \pagesname{} 509 -
  512.

\bibitem[BAR~02]{Barabasi2002}
\textsc{Barabasi A.~L.}, \textit{Linked: The New Science of Networks},
\newblock Perseus Books, 2002.

\bibitem[BOC~06]{Boccaletti2006}
\textsc{Boccaletti S.}, \textsc{Latora V.}, \textsc{Moreno Y.}, \textsc{Chavez
  M.}\andname{}\textsc{Hwang D.}, \guilo{}Complex networks: Structure and
  dynamics\guilf{},
\newblock \textit{Physics reports}, \volumename\ 424, \numbername\ 4, 2006,
  \pagesname{} 175--308, Elsevier.

\bibitem[BOG~03]{Boguna2003}
\textsc{Boguna M.}, \textsc{Pastor-Satorras R.}, \textsc{Diaz-Guilera
  A.}\andname{}\textsc{Arenas A.}, \guilo{}Emergence of clustering,
  correlations, and communities in a social network model\guilf{},
\newblock \textit{Arxiv preprint cond-mat 0309263}, , 2003.

\bibitem[COL~12]{Stattner2012-ANT}
\textsc{Collard M.}, \textsc{Collard P.}\andname{}\textsc{Stattner E.},
  \guilo{}Mobility and information flow: percolation in a multi-agent
  model\guilf{},
\newblock \textit{3rd International Conference on Ambient Systems, Networks and
  Technologies}, , 2012.

\bibitem[DOR~02]{Dorogovtsev2002}
\textsc{Dorogovtsev S.}\andname{}\textsc{Mendes J.}, \guilo{}Evolution of
  networks\guilf{},
\newblock \textit{Adv. Phys}, , 2002.

\bibitem[KOS~06]{Kossinets2006}
\textsc{Kossinets G.}\andname{}\textsc{Watts D.}, \guilo{}Empirical analysis of
  an evolving social network\guilf{},
\newblock \textit{Science}, \volumename\ 311, \numbername\ 5757, 2006,
  \pagesname{} 88--90, American Association for the Advancement of Science.

\bibitem[KUM~07]{Kumpula2007}
\textsc{Kumpula J.}, \textsc{Onnela J.}, \textsc{Saram{\"a}ki J.},
  \textsc{Kaski K.}\andname{}\textsc{Kert{\'e}sz J.}, \guilo{}Emergence of
  communities in weighted networks\guilf{},
\newblock \textit{Physical review letters}, \volumename\ 99, \numbername\ 22,
  2007,  \pagename{} 228701, APS.

\bibitem[LES~08]{Leskovec2008}
\textsc{Leskovec J.}, \textsc{Backstrom L.}, \textsc{Kumar
  R.}\andname{}\textsc{Tomkins A.}, \guilo{}Microscopic evolution of social
  networks\guilf{},
\newblock \Inname{} \textit{Proceedings of the 14th ACM SIGKDD international
  conference on Knowledge discovery and data mining}, ACM, 2008,  \pagesname{}
  462--470.

\bibitem[OPS~11]{Opsahl2011}
\textsc{Opsahl T.}, \guilo{}Triadic closure in two-mode networks: Redefining
  the global and local clustering coefficients\guilf{},
\newblock \textit{Social Networks}, , 2011, Elsevier.

\bibitem[STA~12]{Stattner2012-CHB}
\textsc{Stattner E.}, \textsc{Collard M.}\andname{}\textsc{Vidot N.},
  \guilo{}D2SNet: Dynamics of diffusion and dynamic human behaviour in social
  networks\guilf{},
\newblock \textit{Computers in Human Behavior}, , 2012.

\bibitem[TOI~09]{Toivonen2009}
\textsc{Toivonen R.}, \textsc{Kovanen L.}, \textsc{Kivela M.}, \textsc{Onnela
  J.}, \textsc{Saramaki J.}\andname{}\textsc{Kaski K.}, \guilo{}A comparative
  study of social network models: network evolution models and nodal attribute
  models\guilf{},
\newblock \textit{Social Networks}, \volumename\ 31, 2009.

\bibitem[WAT~98]{Watts1998}
\textsc{Watts D.~J.}\andname{}\textsc{Strogatz S.~H.}, \guilo{}Collective
  dynamics of 'small-world' networks\guilf{},
\newblock \textit{Nature}, \volumename\ 393, 1998,  \pagesname{} 440-42.

\end{thebibliography}
